\documentclass[11pt,a4paper]{iopart}
\pdfoutput=1
\usepackage{graphicx}
\usepackage{dcolumn}
\usepackage{bm}
\usepackage{epstopdf}
\usepackage{xcolor}
\usepackage{epsfig}
\usepackage{float}
\usepackage{stmaryrd}
\usepackage{wasysym}
\graphicspath{{Figures/}}

\begin{document}

\title{Correlations in the shear flow of athermal amorphous solids: A principal component analysis\\}

\author{C\'eline Ruscher and J\"org Rottler}
\address{Department of Physics and Astronomy and Quantum Matter Institute, University of British Columbia, Vancouver BC V6T 1Z1, Canada}

\begin{abstract}
We apply principal component analysis, a method frequently used in image processing and unsupervised machine learning, to characterize particle displacements observed in the steady shear flow of amorphous solids. PCA produces a low-dimensional representation of the data and clearly reveals the dominant features of elastic (i.e.~reversible) and plastic deformation. We show that the principal directions of PCA in the plastic regime correspond to the soft (i.e. zero energy) modes of the elastic propagator that governs the redistribution of shear stress due to localized plastic events. Projections onto these soft modes also correspond to components of the displacement structure factor at the first nonzero wavevectors, in close analogy to PCA results for thermal phase transitions in conserved Ising spin systems. The study showcases the ability of PCA to identify physical observables related to the broken symmetry in a dynamical nonequilibrium transition.
\end{abstract}

\submitto{\JSTAT}
\maketitle

\section{Introduction}

A growing body of research is presently exploring the potential of machine learning (ML) methods as a tool of discovery for new physics \cite{mehta_high-bias_2018}. Much of this work is driven by the expectation that ML might reveal structure and correlations in large data sets obtained either experimentally or numerically that is not directly accessible via conventional analysis. One such method of unsupervised learning that has received significant attention is principal component analysis (PCA). PCA is a statistical dimensionality reduction technique that converts a series of correlated data into a set of uncorrelated values called principal components via a linear transformation. A series of recent papers have argued that PCA is a suitable tool for elucidating phase transitions \cite{wang_discovering_2016,wang_machine_2017,wetzel_unsupervised_2017,hu_discovering_2017,nieuwenburg_learning_2017}. When applied to spin configurations obtained from Monte Carlo simulations of the classical 2D Ising model, the first principal components correctly identify the broken symmetry and the dependence of the global order parameter on temperature. More complex models such as the continuous XY-model or frustrated magnets were also considered. In situations where the underlying Hamiltonian is unknown, PCA or related analysis on raw data may help identifying ordered phases and the transitions or crossovers between them. Connections between PCA and the renormalization group have also been pointed out \cite{bradde_pca_2017,foreman_rg-inspired_2018}.

In the present contribution, we explore the utility of PCA in the analysis of a problem from nonequilibrium statistical physics, namely the slow flow of dense amorphous packings. When subjected to small shear increments,  particles in such materials do not move purely affinely, but exhibit nontrivial correlated residual or nonaffine displacements \cite{tanguy_continuum_2002}. The displacement field exhibits strong rotational character, and their correlations range over a length scale of 20-30 particle diameters and reflect the scale above which the material can be viewed as a homogeneous elastic medium \cite{leonforte_inhomogeneous_2006}. The displacements from individual plastic shear transformations, however, are far more localized and can be thought of as forming at the intersection between (large) vortices      \cite{tanguy_vibration_2015}. When displacements are accumulated over larger strains, the plastic activity focuses particle motion along slip lines or micro shear bands \cite{tanguy_plastic_2006,maloney_evolution_2008,maloney_anisotropic_2009}. It is now well understood that these correlations emerge from a superposition of localized shear transformations whose displacement field has quadrupolar symmetry \cite{maloney_amorphous_2006} and can be modeled as Eshelby inclusions \cite{dasgupta_microscopic_2012,dasgupta_yield_2013}. 

Here we characterize a set of nonaffine displacement fields obtained from molecular simulation of a 2D amorphous solid with PCA. We show that PCA easily differentiates between the dominant features in the elastic and plastic deformation regimes in the athermal quasistatic limit. The PCA principal directions exhibit vortex-like structures in the elastic regime and shear band-like features in the plastic regime, and the latter correspond to null space of the Eshelby propagator that redistributes the shear stress released by plastic events. Driving with finite shear rate reduces the tendency for shear localization. The crossover to homogeneous flow is well captured by an order parameter like quantity formed out of the projections of the displacement field onto the first two PCA components.

\section{Model System and PCA}
\subsection{Simulations}
In order to obtain displacement fields, we study a well-known model system for 2D amorphous materials under simple shear 
i) in the athermal quasistatic limit (AQS) and ii) with molecular dynamics simulations at finite shear rate in the athermal limit \cite{Kob_multiD}. The model glass is a Lennard-Jones (LJ) binary mixture with pairwise interactions described by
\begin{equation}
V_{\alpha\beta}\left(r\right)=4\epsilon_{\alpha\beta}\left[\left(\frac{\sigma_{\alpha\beta}}{r}\right)^{12}-\left(\frac{\sigma_{\alpha\beta}}{r}\right)^{6}\right]
\end{equation}
where $\alpha,\beta=A,\,B$,$\sigma_{AA}=1.0$, $\sigma_{AB}=0.8$,  $\sigma_{BB}=0.88$, $\epsilon_{AA}=1.0$,
$\epsilon_{AB}=1.5$, and $\epsilon_{BB}=0.5$. The potential is truncated
at $r=2.5\sigma_{AA}$ and shifted for continuity. 
We consider $N_{A}=26000$ and $N_{B}=14000$ particles of mass $m=1$ that are placed in a periodic simulation box of dimensions $182\sigma_{AA}\times 182\sigma_{AA}$, corresponding to a density of $1.2$. 
In the following, we use $\sigma_{AA}$ as the unit of length. Working in the $NVT$ ensemble, the system is initially equilibrated at temperature $T= 1.00$ (we recall that for this system $T_g=0.33$ \cite{Kob_multiD}). Then the equilibrated configuration is quenched at cooling rate $dT/dt = 2 \cdot 10^{-3}$. For the AQS protocol, we also perform an energy minimisation after the quench to ensure that the initial configuration corresponds to a minimum of the potential energy landscape. 

Once the starting configuration is obtained we apply the following protocols: 

i) \emph{AQS}: The initially square box is deformed by applying successive strain increments $\delta \gamma = 10^{-5}$. This happens by incrementally tilting the simulation box by a tilt factor $\delta \gamma \times$ box length at each step, so that the simulation box deforms into a parallelogram. The particle positions are then remapped into the new box configuration. After each strain increment, we minimize the potential energy with a conjugate gradient algorithm. 

ii) \emph{Simple shear deformation}:
Simple shear is imposed at rate $\dot{\gamma}$ in the same way as described above, and we integrate the equations of motion in the athermal limit, 
\begin{eqnarray}
\frac{d r_{i}}{dt}  &=  v_{i}\nonumber \\
m\frac{d v_{i}}{dt}  &=  -\sum_{i\neq j}\frac{\partial V\left(r_{ij}\right)}{\partial r_{ij
}}+f_{i}^{D}.
\label{eq_of_motion_MD}
\end{eqnarray}
The dissipative force $f_{i}^{D}$ experienced by particle \emph{i} is computed with a Dissipative Particle Dynamics scheme \cite{groot_dissipative_1997}, i.e. a friction force proportional to the particles' relative velocities. 

In both protocols, nonaffine displacements $\bm{u}(\bm{r})$ are measured in steady state ($> 20 \%$ strain). For particle $j$ the nonaffine displacement is given by \cite{EPL_Goldenberg}
\begin{equation}
u_{j\alpha}=r_{j\alpha} - r_{j\alpha}^{0} - \varepsilon_{\alpha \beta} r^0_{j \beta},
\label{nonaffine_eq}
\end{equation}
where $\varepsilon_{\alpha \beta}$ denotes the strain and Greek letters $\{\alpha, \beta \}$ refer to Cartesian coordinates. The vector $\bm{r}_j^0$ corresponds to the position of the particle at a given strain $\gamma_0$, whereas $\bm{r}_j$ stands for the position of the particle after deformation. Nonaffine displacements are recorded for $n$ configurations and for a given snapshot $i$, the nonaffine displacement vector ${\bm x_i}$ is of dimension $2N$ where $N$ is the total number of particles in the system. The displacement vectors can be grouped into a matrix ${\bf{X}}=(\bm{x}_1,\cdots,\bm{x}_n)$ where the $i^{th}$ row is given by $\bm{x}_i=(x_{i1}, \cdots, x_{i2N})$.

\subsection{PCA}

PCA aims to extract the most important information of a data matrix ${\bf{X}}$ and expresses this information through a matrix of new orthogonal variables ${\bf{Y}}$ called \emph{principal components}  \cite{Abdi_PCA}. PCA assumes that the components of $\bf{Y}$ can be written as a linear combination of the components of $\bf{X}$. Therefore, to preserve most of the information, we look for $\bf{Y} = \bf{X} \bf{W} $ where the elements of $\bm{y}_1,\cdots,\bm{y}_n$ each successively have maximal possible variance.
The data matrix ${\bf{X}}(p, 2N)$ can be preprocessed in such a way that $\bf{X}$ is:\\
(i) centered, $x_{ij}=x_{ij} - (1/p)\sum_{j=1}^p x_{ij}$, where $p$ is the total number of principal directions \\
(ii) normalised, $|| \bm{x}_i ||=1$ \\
Maximising the variance of $\bf{Y}$ is equivalent to maximising the quadratic form $\bf{W}^{\rm{T}} \bf{C}_{\bf{X}} \bf{W}$, where $\bf{C}_{\bf{X}}=\bf{X}^{\rm{T}} \bf{X}$ is the correlation matrix, with the restriction that ${\bf{W}^{\rm{T}}\bf{W}} = {\bf{I}}_p$. 
The method of Lagrange multipliers shows that this is achieved by finding the eigenvectors of $ \bf{C}_{\bf{X}}$ \cite{Joliffe_Book, Joliffe_Article},
\begin{equation}
\bf{W}^{\rm{T}} \bf{C}_{\bf{X}} \bf{W}	= \bf{\Lambda}^2 \hspace{0.2cm} \rm{with}  \hspace{0.2cm} {\bf{W}^{\rm{T}}\bf{W}} = {\bf{I}}_p
\end{equation}
where the eigenvalues $\lambda_i =\sqrt{\Lambda^2_i} $ of the correlation matrix $\bf{C}_{\bf{X}}$ are ordered in descending order. The normalized vectors $(\bm{w}_1 \cdots \bm{w}_n)$ define a new orthonormal basis, where $\bm{w}_1$ is the direction with the maximal variance, $\bm{w}_2$ the direction with the second largest variance, etc. 
 
PCA implies dimensionality reduction, which means that only a small number of eigenvectors carry most of the information $(70\%-80\%)$ of the original data \cite{Joliffe_Book, Joliffe_Article}. The normalised eigenvalue $\tilde{\lambda}_i$, 
\begin{equation}
\tilde{\lambda_i}=\frac{\lambda_i}{\sum_i\lambda_i}
\label{explained_variance_ratio},
\end{equation}
also called explained variance ratio, quantifies the relative importance of each eigenvalue $\lambda_i$ (and the associated eigenvector $\bm{w}_i$). In what follows, we will also be interested in the quantified principal components, which correspond to the averaged projections onto a given eigenvector \cite{hu_discovering_2017}, 
\begin{equation}
\label{qpc-eq}
\langle |y_{\ell}|^2 \rangle= \frac{1}{n} \sum_i |{\bf x}_i\cdot{\bf w}_{\ell}|^2.
\end{equation}

The quality of PCA can measured by the reconstruction error, which quantifies the discrepancy between the data matrix and the projected data matrix obtained from the knowledge of the principal components ${\bf{Y}}$ and the eigenvectors ${\bf{W}}$:

\begin{equation}
{\rm error} = \langle || {\bf{X}} - {\bf{Y} \bf{W}^{\rm{T}}} ||^2 \rangle	
\label{error}
\end{equation}

PCA is implemented via the decomposition module in the {\tt scikit-learn} python library \cite{scikit-learn}. 

\begin{figure}[t]
\begin{center}
\includegraphics[scale=0.45]{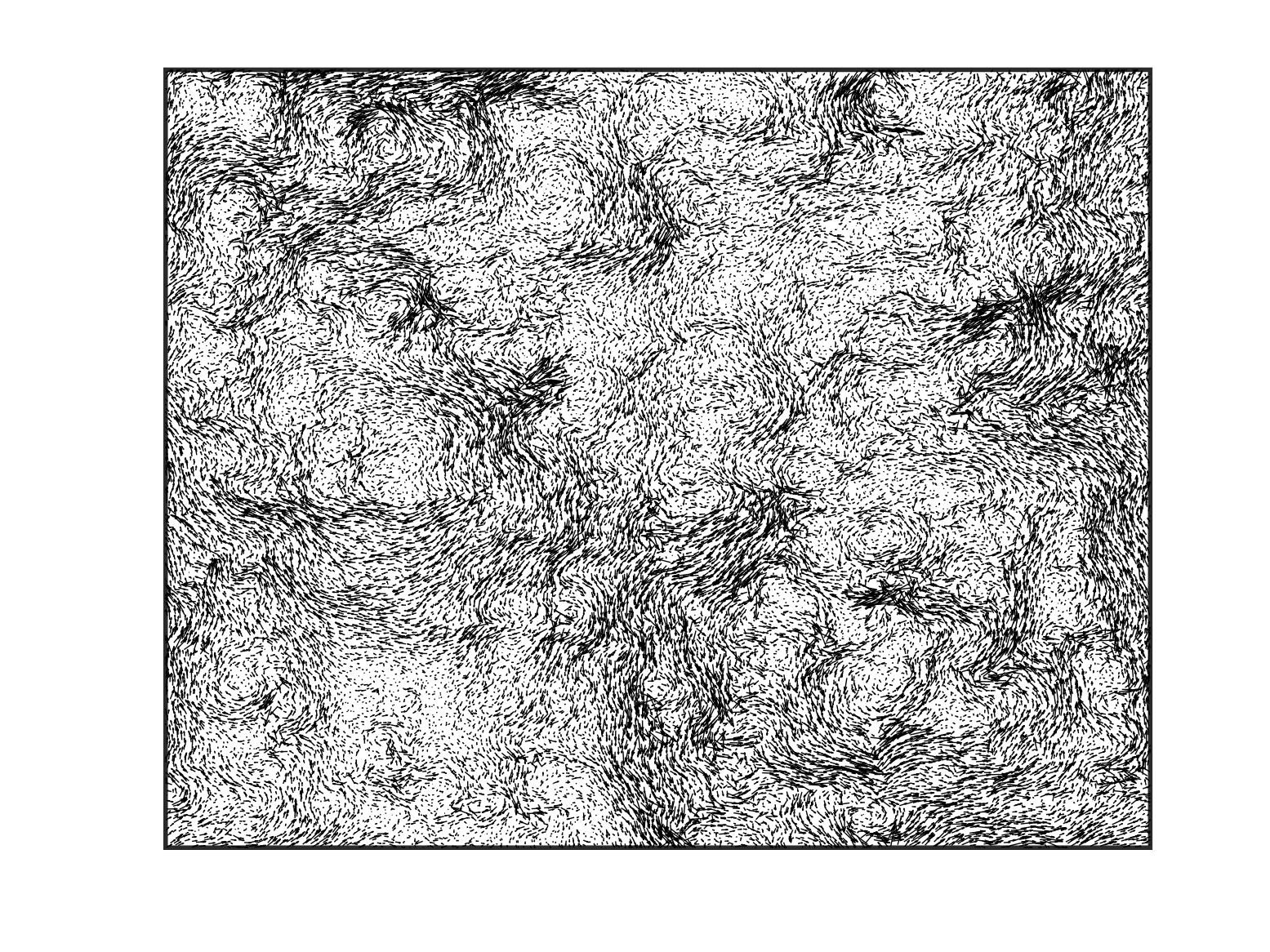}\\
\includegraphics[scale=0.45]{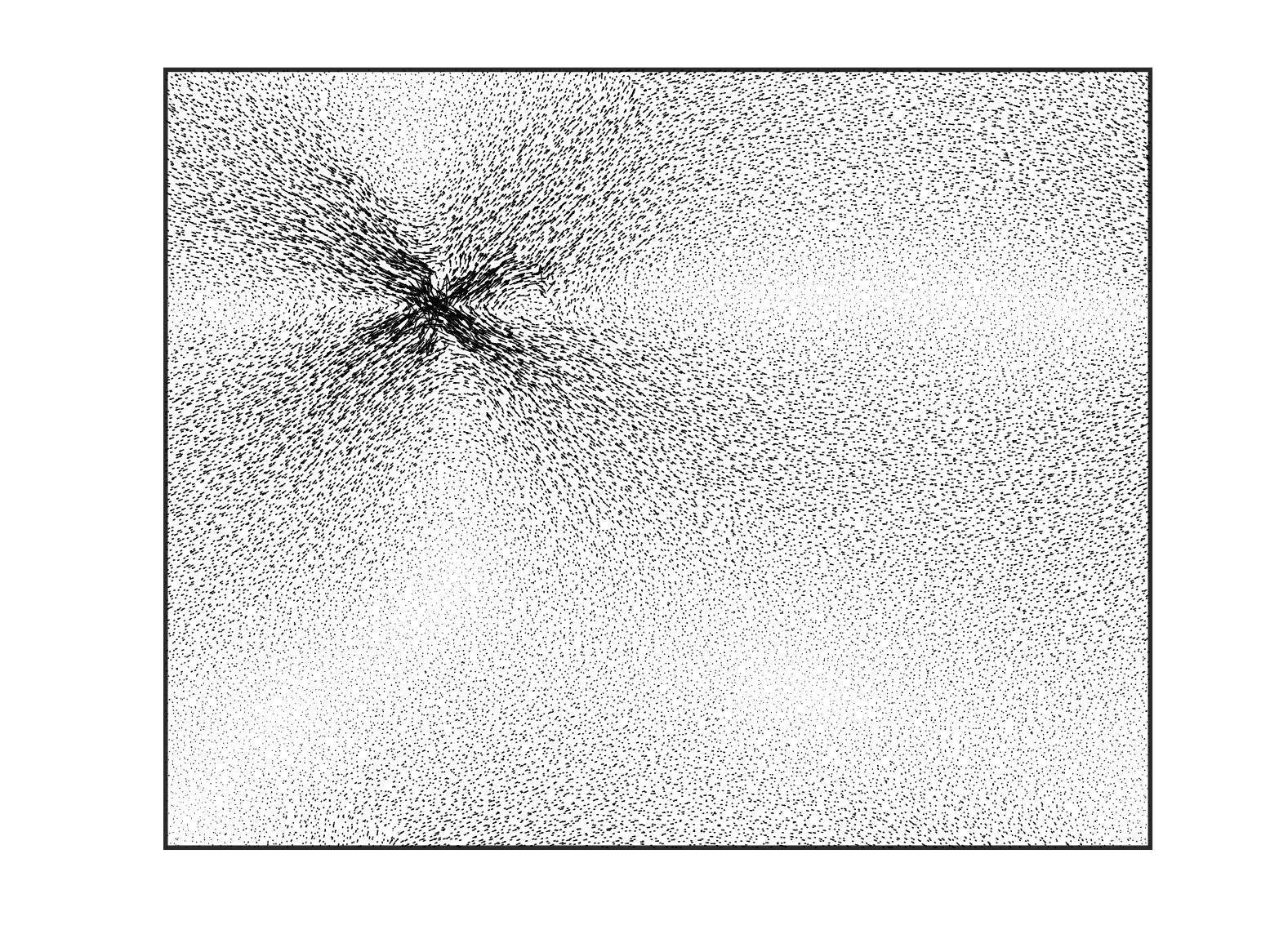}
\end{center}
\caption{Nonaffine displacement fields observed with the AQS protocol in the elastic (top) and plastic branches (bottom). }
\label{displacement_fields_AQS}
\end{figure}

\section{Elastic vs. plastic displacements}
In the AQS protocol, the stress-strain curve can be clearly decomposed into elastic branches that are punctuated by irreversible plastic events. 
Plastic events are associated with stress release and correspond therefore to drops in the stress-strain curve. Nonaffine displacements are recorded during the stress drops and also in the elastic regime of duration $\Delta \gamma$ that precedes the plastic event. 
To be sure that we are probing reversible dynamics in the elastic regime, a reverse strain step of size $-\Delta \gamma$ is systematically applied. By doing so, we find that $\sim 10 \% $ of the elastic branches exhibit irreversible rearrangements. These branches were discarded for the analysis as we aim to probe pure reversible transformation. We record 5000 events in total. Typical nonaffine displacements fields are shown in Figure \ref{displacement_fields_AQS} for both regimes. As reported in many previous works \cite{tanguy_plastic_2006, tanguy_continuum_2002, maloney_amorphous_2006, tanguy_vibration_2015}, localized large displacements with distinct quadrupolar symmetry are associated with the plastic regime, whereas the elastic regime is characterized by extended vortices.

\begin{figure}[t]
\begin{center}
\includegraphics[scale=0.5]{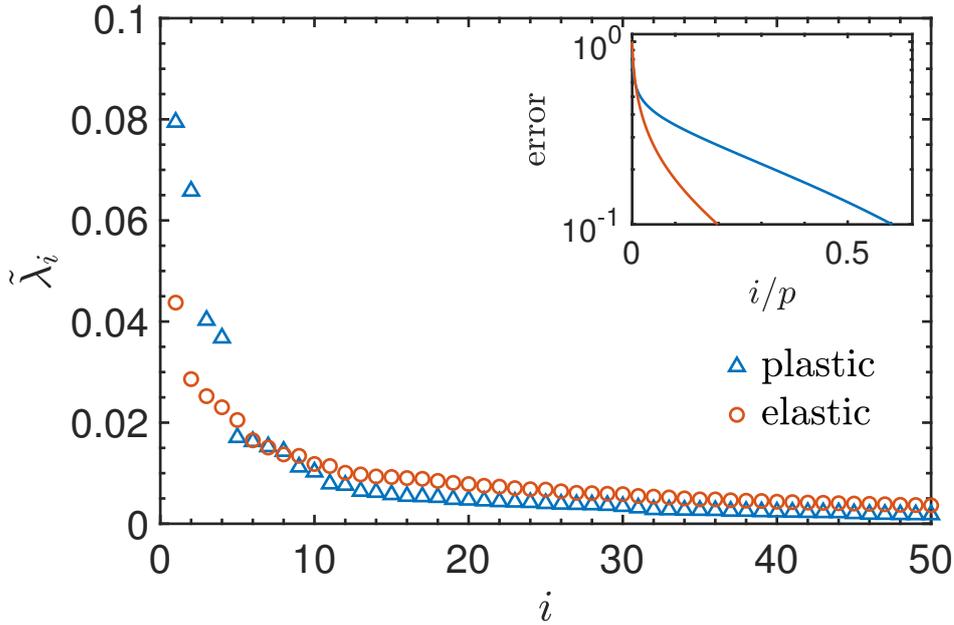}
\caption{Main panel: Explained variance ratios eq.~(\ref{explained_variance_ratio}) for plastic (blue $\triangle$) and elastic (orange $\fullmoon$) regimes.  Inset: Reconstruction error as a function of the number of principal directions $i/p$ normalised by the total number $p=5000$ of principal directions. }
\label{explained_variance_ratios_AQS}
\end{center}
\end{figure}

PCA applied to the nonaffine displacements of both elastic and plastic branches reveals that the information is distributed among a relatively large number of principal components. 
Indeed, the first explained variance ratios $\tilde{\lambda}_i$, shown in Figure \ref{explained_variance_ratios_AQS}, are relatively small (less than $10 \%$). Moreover, when we compute the error defined in eq.~(\ref{error}) to estimate the difference between the original data matrix and its reconstruction, we find that the fraction of principal components required to reduce the error to $90\%$ of its original value is of the order $20\%$ for the elastic regime, whereas it reaches $60\%$ for the plastic regime.

This difference between the two regimes might be explained by the strong localization of the nonaffine displacement field in the plastic case. Successive plastic events may occur at different places and in the simulation box and posess different orientations. By contrast, in the elastic branches the vortices are more extended in space, and consecutive snapshots of elastic branches are more likely to share similarities. As PCA aims to identify the similarites in different snapshots, a larger number of directions may be needed to capture the information about the smaller features in the plastic regime than about the larger displacement patterns in the elastic regime.


\begin{figure}[H]
\begin{center}
\includegraphics[scale=0.78]{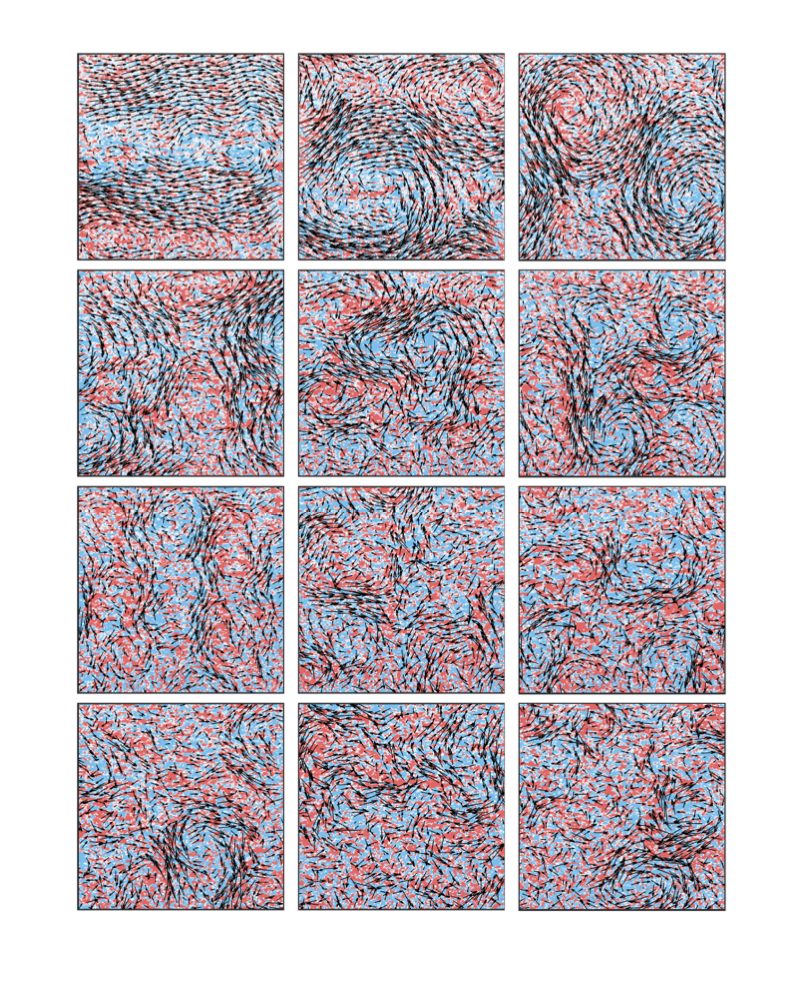}
\end{center}
\caption{The first 12 first eigenvectors of the covariance matrix associated with the elastic branches (rowwise top to bottom).  The colors indicate the value of the vorticity $(\partial u_x / \partial y - \partial u_y / \partial x) \in [-5 \cdot 10^{-4} ; 5 \cdot 10^{-4}]$ (blue to red).  }
\label{elastic_eigenvectors_AQS}

\end{figure}

\begin{figure}[H]
\begin{center}
\includegraphics[scale=0.78]{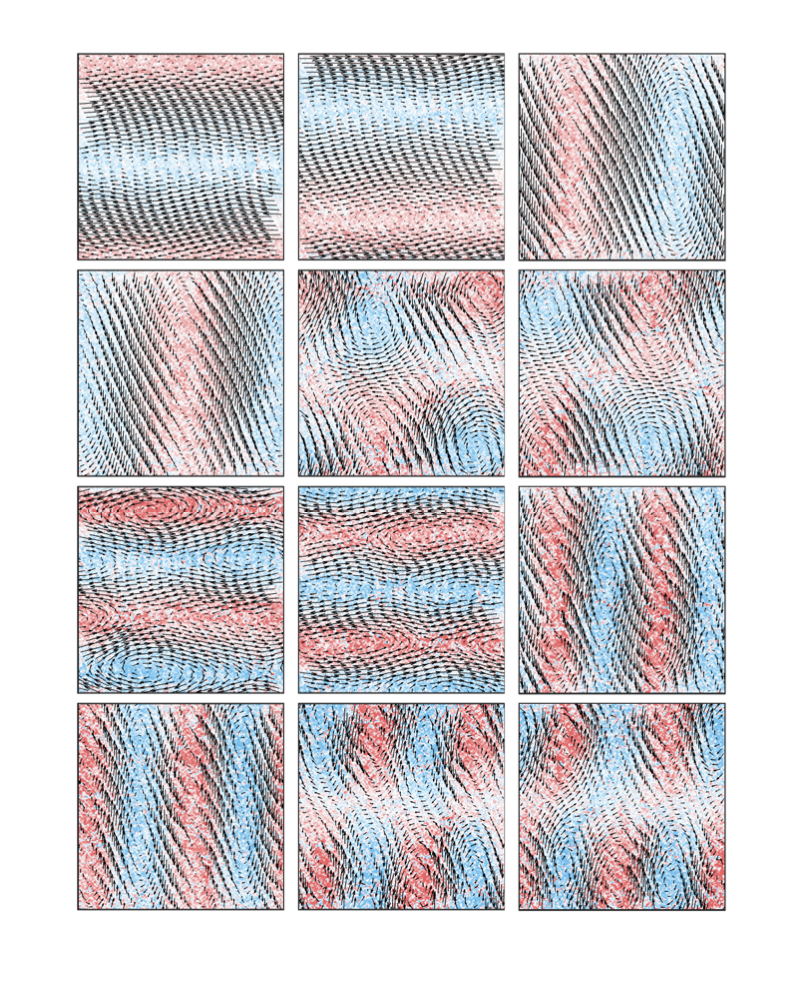}
\end{center}
\caption{The first 12 first eigenvectors of the covariance matrix associated with the plastic branches (rowwise top to bottom).  Colors indicate the value of the vorticity $(\partial u_x / \partial y - \partial u_y / \partial x) \in [-5 \cdot 10^{-4} ; 5 \cdot 10^{-4}]$ (blue to red). }
\label{plastic_eigenvectors_AQS}

\end{figure}


Despite the absence of strongly predominant directions, the first explained variance ratios reveal a marked difference between the elastic and plastic branches. In the elastic case, only the first eigenvalue is slightly larger than the subsequent ones, whereas in the plastic case, the first four eigenvalues are signficantly larger and exhibit a two-step pattern. Inspection of the associated eigenvectors in Figures \ref{elastic_eigenvectors_AQS} and \ref{plastic_eigenvectors_AQS} unveils the reason for this behaviour. In the elastic regime, the first eigenvector is reminiscent of horizontal shear bands. This behaviour results from the alignment of large vortices \cite{tanguy_plastic_2006}. The 2nd to 5th eigenvectors show large vortices that resemble the displacement fields associated with low frequency modes \cite{Jojo, tanguy_continuum_2002}. Higher order eigenvectors appear to be a combination of shear bands or vortices. Therefore, the ranked eigenvalues of the elastic regime shown in Fig.~\ref{explained_variance_ratios_AQS} exhibit a smooth decrease with increasing index, consistent with the absence of a dominant pattern.

Very different behaviour is observed in the plastic regime, see Fig.~\ref{plastic_eigenvectors_AQS}. Here, the first two eigenvectors evidently show a purely horizontal displacement field modulated by $\sin(2\pi y/L)$ and $\cos(2\pi y/L)$, while the third and fourth show purely vertical displacements modulated by $\sin(2\pi x/L)$ and $\cos(2\pi x/L)$. The higher order modes are combinations of superpositions of the above (5th and 6th eigenvectors) and modulations at twice the frequency. Thus the first four ranked eigenvalues of the plastic regime in Fig.~\ref{explained_variance_ratios_AQS} are larger than those in the elastic regime and appear structured into groups of two. \\

What is the significance of this PCA decomposition? In order to understand its physical origin, we recall that plastic flow in amorphous solids is mediated by localized shear transformations. The shear stress in a 2D elastic medium responding to such a transformation at the origin is proportional to the {\em Eshelby propagator} 
\begin{equation}
G({\bf r})= \frac{\cos(4\theta)}{r^2}
\end{equation} 
in polar coordinates. Its quadrupolar symmetry (and resulting alternating sign) is responsible for many important properties of the yielding transition. In Fourier space, this propagator can be written as 
\begin{equation}
\tilde{G}({\bf q})=-\frac{4q_x^2q_y^2}{q^4},
\end{equation} 
where ${\bf q}$ is a 2D wavevector and $q_x,q_y=2\pi n/L$ in a periodic system of size $L$. The elastic shear stress from a plastic shear strain field $\varepsilon_{pl}({\bf q})$ is then proportional to $\tilde{G}({\bf q})\varepsilon_{pl}({\bf q})$. In this representation, it can be seen that the eigenmodes (plane waves) of the progagator that cost no energy satisfy $\tilde{G}({\bf q})\varepsilon_{pl}({\bf q})=0$, i.e. they are the nullspace of the propagator. As pointed out by Tyukodi {\it et al.} \cite{Tyukodi} these soft modes correspond to eigenmodes characterized by a null eigenvalue (thus either $q_x=0$ or $q_y=0$) and lead to horizontal or vertical shear bands that produce no energy. This physics emerges precisely in the form of the first four eigenvectors of the PCA transformation. Indeed we expect shear bands to develop in the horizontal direction where shear is applied, and this preference is reflected in the larger eigenvalues of the first two eigenvectors.

\begin{figure}[t]
\begin{center}
\includegraphics[scale=0.5]{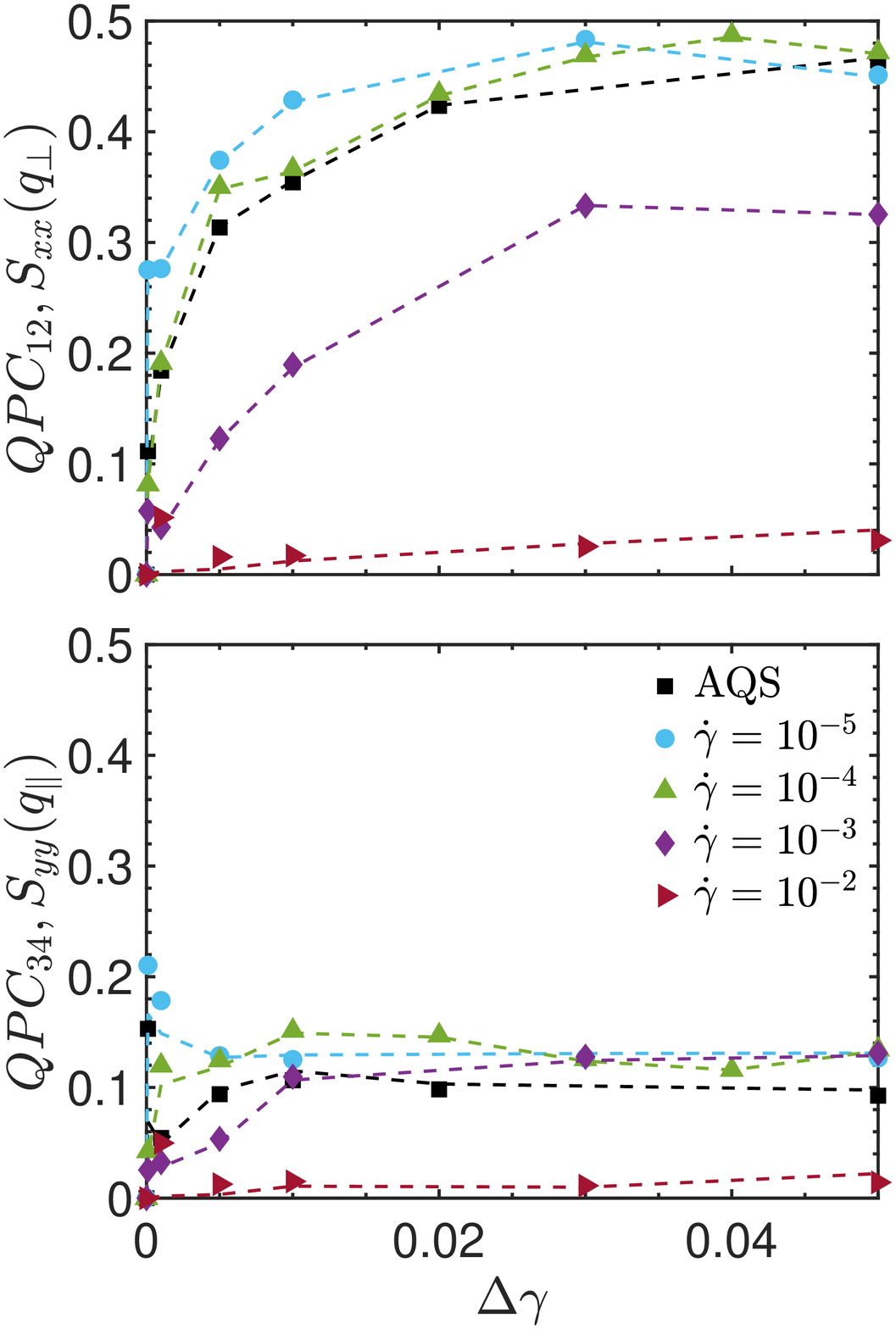}
\end{center}
\caption{Quantified principal components eq.~(\ref{op-eq}) (symbols) and components of the displacement structure factor eq.~(\ref{S-eq}) (dashed lines) as a function of strain interval $\Delta \gamma$. }
\label{fig-qpc}
\end{figure}

\section{Effect of strain accumulation and shear rate}

In the previous section, we focussed our analysis on separating elastic and plastic behavior. We now accumulate nonaffine displacement fields over a fixed strain interval $\Delta \gamma \in [10^{-5}; 10^{-2}]$, and perform PCA on the entire dataset that contains 500 snapshots for each value of $\Delta \gamma$.  We also investigate how PCA views the displacement fields when the glass is flowing at finite shear rate $\dot{\gamma}$. To this end, we switch from the AQS protocol to the simple shear protocol (see Section 2) and collect also 500 samples for each value of $\Delta \gamma$ at four different strain rates $\dot{\gamma}$.
As $\Delta \gamma$ increases, we expect that more and more plastic events are being sampled and self-organize into shear bands.

In order to quantify this transition from elastic reponse to plastic flow with increasing strain interval, a suitable observable needs to be defined. The structuring of the first four explained variance ratios observed in Fig.~\ref{explained_variance_ratios_AQS} suggests to consider a separate superposition of the horizontal and vertical principal components,
\begin{equation}
\label{op-eq}
QPC_{12}(\Delta \gamma)=\sum_{\ell=1}^2\langle |y_{\ell}|^2 \rangle, \qquad QPC_{34}(\Delta \gamma)=\sum_{\ell=3}^4\langle |y_{\ell}|^2 \rangle
\end{equation}
where the $\langle |y_{\ell}|^2 \rangle$ are defined in eq.~(\ref{qpc-eq}) and include observations for a given value of $\Delta \gamma$. In Fig.~\ref{fig-qpc} we see that these two quantities rise rapidly with accumulated strain and saturate at $\Delta\gamma\approx 0.01$. This value is about 20 times larger than the typical elastic strain interval interval $\Delta \gamma_{el}=5\cdot 10^{-4}$ measured in the AQS limit, suggesting that roughly 20 plastic events are required for the formation of a shear band. The average spacing of these events in our system is then approximately equal to $9 \sigma_{AA}$, which is comparable to the typical size of a shear transformation \cite{Nicolas}.
Therefore, $QPC_{12}$ and $QPC_{34}$ describe a transition from elastic to plastic flow, and the preference of horizontal over vertical shear bands is reflected by $QPC_{12}>QPC_{34}$.

\begin{figure}[t]
\begin{center}
\includegraphics[scale=0.5]{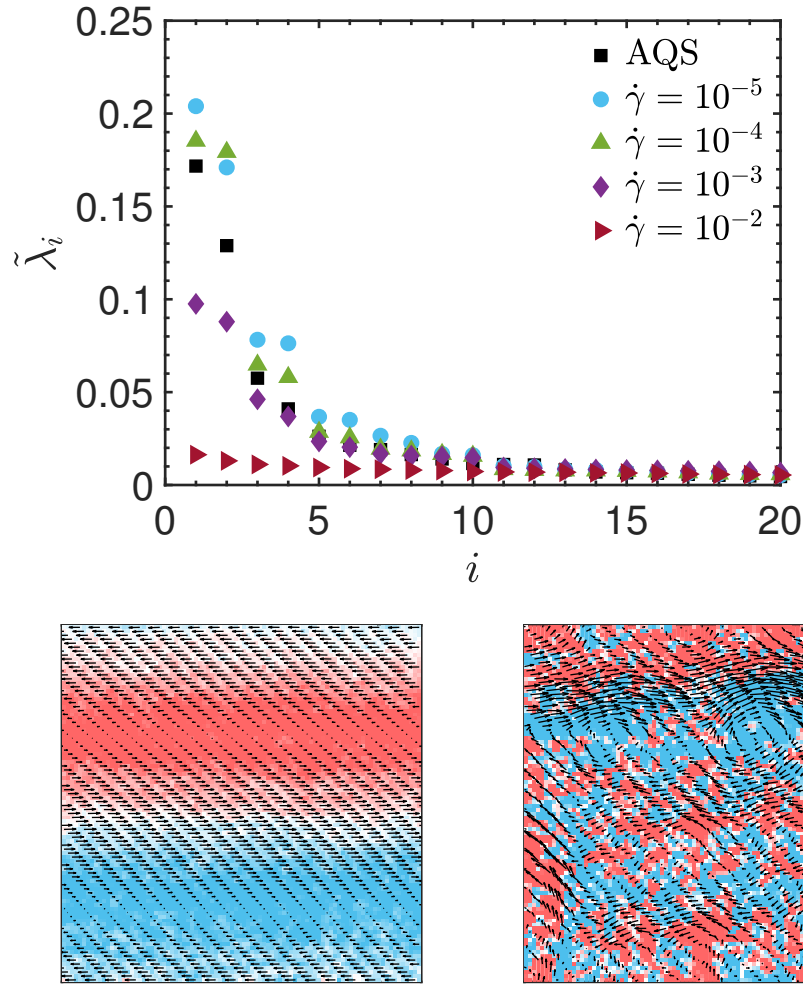}
\caption{Effect of accumulating events in AQS and finite shear simulations on the number of principal directions. Bottom graphs show the first eigenvector for  $\dot{\gamma}=10^{-5}$ (left) and $\dot{\gamma}= 10^{-2}$ (right). Colors indicate the value of the vorticity $(\partial u_x / \partial y - \partial u_y / \partial x) \in [-4 \cdot 10^{-4} ; 4 \cdot 10^{-4}]$ (blue to red).   }
\label{effect_shear_rates_eigval}
\end{center}
\end{figure}

A useful physical interpretation of these observables can be obtained by recalling that the quantified principal components are just averaged projections of the sets of nonaffine displacements $\boldsymbol{u}(\boldsymbol{r})$ onto a particular eigenvector. Since the first four eigenvectors are particularly simple and just represent sinusoidally modulated displacements that point either in the x- or y-direction, the quantified principal components are proportional to components of the (static) displacement vector structure factor \cite{wang_discovering_2016}
\begin{equation}
\label{S-eq}
S_{\alpha\beta}({\bf q})=\langle \tilde{u}_\alpha({\bf q})\tilde{u}_\beta(-{\bf q})\rangle,
\end{equation}     
where $\tilde{{\bf u}}({\bf q})=\frac{1}{\sqrt{N}}\sum_{i=1}^N\exp({ \rm i}{\bf q} \cdot {\bf r}_i){\bf u}({\bf r}_i)/||{\bf u}({\bf r}_i)||$ are the (normalized) nonaffine displacements in reciprocal space. Specifically, selecting the first nonzero wavevectors ${\bf q}_{||}=(2\pi/L,0)$ and ${\bf q}_{\perp}=(0,2\pi/L)$ parallel and perpendicular to the shear direction should reproduce $QPC_{12}$ as $S_{xx}({\bf q}_{\perp})$ and $QPC_{34}$ as $S_{yy}({\bf q}_{||})$. Indeed  Fig.~\ref{fig-qpc} indicates near perfect agreement between these quantities. The appearance of plane wave like eigenvectors in PCA is understood to be a more general feature related to the fact that the largest eigenvalues in PCA are probing the low frequency modes developing when $\Delta \gamma$ increases. As a result, PCA emphasizes the redundant information related to the strong spatial correlation of the nonaffine displacement field \cite{wang_discovering_2016, novembre2008interpreting}.  



Direct comparison of finite shear rate simulations with the AQS results in Fig.~\ref{fig-qpc} shows that the quantified principal compoments at shear rate $\dot{\gamma}=10^{-5}$ are larger than the AQS results, which themselves are more comparable to those obtained for $\dot{\gamma}=10^{-4}$. This trend is also visible in the evolution of eigenvalues shown in Fig.~\ref{effect_shear_rates_eigval} and is surprising given that the AQS protocol usually represents the zero shear rate limit. It seems to indicate that the details of the numerical protocol have some impact on the way plastic events accumulate. While the $QPC_{12}$-values of the two lower shear rates and the AQS limit all follow the same trend, a decrease is observed once $\dot{\gamma}\ge 10^{-3}$, which signals that shear bands are less pronounced at higher driving. The $QPC_{34}$-values remain consistently lower, reminding us that the plastic events organize predominantly in horizontal bands and not in vertical ones. At the highest driving rate considered, $\dot{\gamma}= 10^{-2}$, we find $QPC_{12}=QPC_{34}$, which means that at this fast shear rate the flow is entirely homogeneous. These observations are also supported by the behavior of the explained variance ratios shown in Fig.~\ref{effect_shear_rates_eigval}. For  $ \dot{\gamma} \le 10^{-3}$ they exhibit the pattern associated with the presence of horizontal and vertical bands. By contrast, for $ \dot{\gamma} \le 10^{-2}$, the very low values of the eigenvalues and the nearly flat variation with the increasing number of directions imply the absence of a specific pattern and an equivalent contribution of the directions. The associated first eigenvectors shown in the bottom part of Fig.~\ref{effect_shear_rates_eigval} highlight the difference between localized and homogeneous flow for the slowest and fastest shear rates, respectively.

\section{Discussion and Conclusion}
Previous applications of PCA to disordered materials have focused mainly on the covariance matrix of particle {\em positions}, where the principal components can be interpreted as vibrational normal modes. This analysis revealed for instance that when a granular glass approaches the jamming transition from above, a smaller and smaller amount of large, collective modes concentrate the dynamics for longer and longer times \cite{Brito}. These soft vibrational models facilitate rearrangements in jammed materials. Here, we show that PCA can also play a productive role in the analysis of nonaffine {\em displacements} in steadily sheared amorphous materials. PCA emphasizes an elastic to plastic transition that depends on the size of the strain interval over which the flow is observed by distinguishing different patterns that are exclusively associated with either regime. For plastic events, PCA robustly identifies the principal symmetry of shear deformation in the form of horizontal and vertical shear bands that originates in the soft modes of the quadrupolar elastic interaction produced by a shear transformation \cite{Tyukodi}. 

An interesting analogy can be drawn with equilibrium phase transitions in Ising ferromagnets on a 2D lattice \cite{wang_discovering_2016}. For these systems, PCA was shown to successfully identify the correct physically relevant order parameter. For nonconserved dynamics, this order parameter is the magnitude of the total (uniform) magnetization, or equivalently the $\boldsymbol{q}=0$ - value of the spin structure factor. When the constraint of conserved magnetization $(\sum_i \sigma_i=0)$ is imposed (when modeling for instance lattice gases), the ground state consists instead of two either horizontal or vertical domains to minimize the domain wall energy. PCA analysis of this system now reveals four dominant eigenvalues and plane-wave like eigenvectors \cite{wang_discovering_2016}. As in the present case, the relevant order parameter is the spin structure factor evaluated at the first nonzero wavevectors. This is no coincidence if we recall that the nonaffine displacement field is also subject to the same constraint, because its integral over the periodic domain must vanish \cite{maloney_evolution_2008}. 

Therefore, if we consider the elastic (vortex-dominated) and plastic (shear-banded) response as the nonequilibrium counterpart to the disordered and ordered phases of the Ising ferromagnet, then our principal components identify the lower-symmetry configuration, and the sum of the first two quantified principal component serves as an order parameter that is equivalent to the smallest nonzero Fourier mode in the displacement structure factor. To our knowledge, this order parameter has not been proposed in the context of plasticity in amorphous solids. Moreover, PCA distiguishes between well structured deformation patterns at low shear rate and the more homogeneous, fluid like behaviour at high shear rate via a reduction of the expected variance ratio and quantified principal components. 

This direct geometric interpretation makes it attractive to consider PCA as an alternative tool to analyse data about which little information is available and for which no obvious order parameter can be identified. PCA could have several useful other applications in the mechanics of amorphous solids, for instance in characterizing the brittle-ductile transition as a function of degree of annealing of the glass \cite{Ozawa}. It could be interesting to see if PCA can characterize the nature of this transition by distinguishing between homogeneous and localized flow. Another area of interest might lie in the comparison of strain correlations between particle scale and mesoscopic simulations, which has so far only been done based on conventional correlation functions \cite{Nicolas}. Comparing directly principal components between the two models could lead to improved benchmarks and calibration procedures.

\section*{Acknowledgements}
We gratefully acknowledge support from the Discovery Grant program of the Natural Sciences and Engineering Research Council of Canada.

\end{document}